\documentstyle[prl,aps,floats]{revtex}
\begin{document}
%
%
\input epsf
\renewcommand{\topfraction}{1}
\preprint{hep-ph/}
\title{The (in)stability of global monopoles revisited}
\author{Ana Ach\'{u}carro$^{1,2}$ and Jon Urrestilla$^1$}  
\address{$^1$Department of Theoretical Physics, UPV-EHU, Bilbao, Spain\\
$^2$Institute for Theoretical Physics, University of Groningen, The
Netherlands}
\date{\today}
\maketitle
\begin{abstract}
We analyse the stability of global O(3) monopoles in
 the infinite cut-off (or scalar mass) limit. We obtain the perturbation
 equations and  
 prove that the
 spherically symmetric solution is classically {\it stable} (or {\it neutrally stable}) to 
axially symmetric, square integrable or power-law decay 
 perturbations. Moreover we show that, in spite of 
the existence of a
conserved topological charge, the energy barrier between the monopole
and the vacuum is {\it finite} even in the limit where the cut-off 
is taken to infinity. This feature is specific of global monopoles and independent of the details of
the scalar potential.   
\end{abstract}

\def\rd{{\rm{d}}}
\def\d{\partial}
\def\half{{1 \over 2}}
\def\quarter{{1 \over 4}}
\def\thetabar{\bar{\theta}}
\def\sech{{\rm sech}}
\def\tanh{{\rm tanh}}
\def\simleq{\; \raise0.3ex\hbox{$<$\kern-0.75em
      \raise-1.1ex\hbox{$\sim$}}\; }
\def\simgeq{\; \raise0.3ex\hbox{$>$\kern-0.75em
      \raise-1.1ex\hbox{$\sim$}}\; }  

\section{Introduction}      
Global monopoles have been investigated for years as possible seeds
for structure formation in the Universe \cite{VS,MB00}.  Although
they appear to be ruled out by the latest
cosmological data \cite{DKM99}, their appearance in condensed matter
--and other-- systems and their peculiar properties make them worthy of
investigation.  These objects have divergent energy, due to the slow
fall-off of angular gradients in the fields, which has to be cut-off
at a certain distance  R  (in practice, the
distance to the nearest monopole or antimonopole) and has two important consequences, in
particular for cosmology. First, the evolution of a network of
global monopoles is very different from that of gauged monopoles, as
long-range interactions enhance annihilation to the extent of
eliminating the overabundance problem altogether \cite{VS}. Second,
 their gravitational properties
include a deficit solid angle \cite{BV}, which makes them rather exotic.

The stability of global $O(3)$ monopoles has been the subject of some
debate in the literature \cite{G,BR,P}. In this paper we try to settle
the issue by: a) analysing the axial perturbation equations in the limit
where the cut-off is taken to infinity, and b) proving that the energy
barrier between the monopole and the vacuum (meaning the {\it extra}
energy required by the monopole to reach an unstable configuration
that decays to the vacuum) is {\it finite}. It is somewhat surprising
for different topological sectors to be separated by finite energy
barriers, but in this case it is a consequence of the scale invariance
of gradient energy on two dimensional surfaces ($r=$ constant), and
therefore independent of the details of the scalar potential.
\section{The Model}
We consider the simplest model that gives rise to global monopoles,
the $O(3)$ model with lagrangian:
\begin{equation}
L = {1 \over 2} \d_{\mu}\Phi^a \d^{\mu}\Phi^a -
        {1 \over 4} \lambda (|\Phi|^2-\eta^2)^2 \   \qquad  a=1,2,3. 
\label{lagrangian}
\end{equation}
$ \Phi^a $ is a scalar triplet, $|\Phi|\equiv \sqrt{\Phi^a \Phi^a}$ 
and $\mu=0,1,2,3$.  The $O(3)$
 symmetry is spontaneously broken to $O(2)$, leading to two
 Goldstone bosons and one scalar excitation with mass $m_s=\sqrt{2
 \lambda} \eta$.  The set of ground states is the two-sphere
 $|\Phi|=\eta$ and, since $\pi_2(S^2) = {\rm \bf Z}$, there are field
 configurations with non-trivial topological charge. One such
 configuration with unit winding is the 
 spherically  symmetric monopole,
\begin{equation}
        \Phi^a = \eta f(r) {x^a \over r}\   \qquad ,
\label{mon}
\end{equation}
where $f(0)=0$ and $f(r\to\infty)=1$. Its
 asymptotic behaviour is $f(r\to0) \sim \alpha r$,
$\alpha \approx 0.5$  and $f(r\to\infty) \sim 1 - 1/r^2 - 3/2r^4$, as can 
be seen from the e.o.m. of $f(r)$ 
\begin{equation}
f''+\frac{2}{r}f'-\frac{2}{r^2}f-f\left(f^2-1\right)=0\qquad .
\label{eom}
\end{equation}

The two parameters ($\eta,\lambda$) appearing in the lagrangian can be
absorbed by the rescaling $ \Phi^a \to \tilde{\Phi}^a =
{\Phi^a / \eta}, \quad x^\mu \to \tilde x^\mu =\sqrt{\lambda
\eta^2}x^\mu , $ which amounts to choosing $\eta$ as the
unit of energy and the inverse scalar mass as the unit of length (up
to a numerical factor). Note, however, that the energy of a
configuration with non-trivial winding such as (\ref{mon}) is (linearly)
divergent with radius, due to the slow fall-off of angular gradients
and has to be cut off at 
$r=R$, say. Unlike $\eta $ and $\lambda$, the (rescaled) cut-off is
an important parameter which could affect the dynamics of solutions with
non-trivial topology. Dropping tildes: 
\begin{equation} E = \int_0^{{R}}{1 \over 2}
\d_{\mu}{\Phi}^a \d^{\mu}{\Phi}^a + {1 \over 4} (|{\Phi}|^2-1)^2\ \qquad .
\label{rescaled-energy}
\end{equation}
Since the energy diverges, Derrick's theorem does not apply
in this case, and in \cite{P} it was shown that the global
monopole is stable towards radial rescalings.  On the other hand the
question of stability with respect to angular perturbations has led to
some discussion in the literature after Goldhaber \cite{G} pointed out 
that the ansatz
\begin{eqnarray}
\Phi^1& = &F(r,\theta) \sin\bar{\theta}(r,\theta) \cos\varphi \nonumber\\
\Phi^2& = &F(r,\theta) \sin\bar{\theta}(r,\theta) \sin\varphi \nonumber\\
\Phi^3& = &F(r,\theta) \cos\bar{\theta}(r,\theta)\  \ , 
\label{ansatz}
\end{eqnarray}
which describes axially symmetric deformations of the
spherical monopole (\ref{mon}), leads to the
following expression for the energy 
after a change of variables $y=\ln\tan({\theta /2}$):
\begin{eqnarray}
&E = \int_0^{2 \pi} \rd \varphi \int_{-\infty}^{\infty} \rd y \int_0^R \rd r\,
 {1
 \over 2}\, \left[ \rho_1 + r^2 \sech^2(y) \rho_2\right] \nonumber\\
&\rho_1 = F_y^2+F^2 [\sin^2({\bar \theta}) + ({\bar \theta}_y)^2] \nonumber\\
&\rho_2 = F_r^2+ F^2 ({\bar \theta}_r)^2 + {1 \over 2}
(F^2 -1 )^2 
\label{energy}
\end{eqnarray}
with $F_r \equiv \d_r F$, etc. 
The term in brackets in $\rho_1$
is identical to
the energy of the sine-Gordon soliton, so
translational invariance in $y$ implies that configurations with 
\begin{equation} 
F(r,y)=f(r),\quad \tan\left({\bar \theta}(r,y)/ 2\right)=
e^{y + \xi}, \quad  \xi={\rm const}
\label{globalshift}
\end{equation} 
have the same energy as (\ref{mon}) (which
corresponds to $\xi =0$). On a given $r =$const. shell,
 the effect of taking $\xi \to \infty$ 
is to concentrate the angular gradients in an arbitrarily
small region around the north pole. When the
gradient energy is inside a region of size comparable to the
inverse scalar mass, it is energetically favourable to ``undo the
knot'' by reducing the modulus of the scalar field to zero and
climbing over the top of the mexican hat potential. Unwinding is
 estimated to occur at a critical value of $\xi$ (say $\xi_0$)
 whose dependence with $r$ far from the core is logarithmic, $\xi_0
\approx \ln r + {\rm const.}$ Numerical simulation on individual shells
closer to the core gives $\xi_0 \approx a_0 + b_0 \ln r$ with slowly varying 
${\it b}_0\approx 1 $ and  $ {\it a}_0 \approx -1.3$.  Unwinding is
expected if $\xi(r) > \xi_0(r)$.

\section{Estimation of the energy barrier}

\begin{figure*}[t]
\centering
\leavevmode\epsfysize=10cm \epsfbox{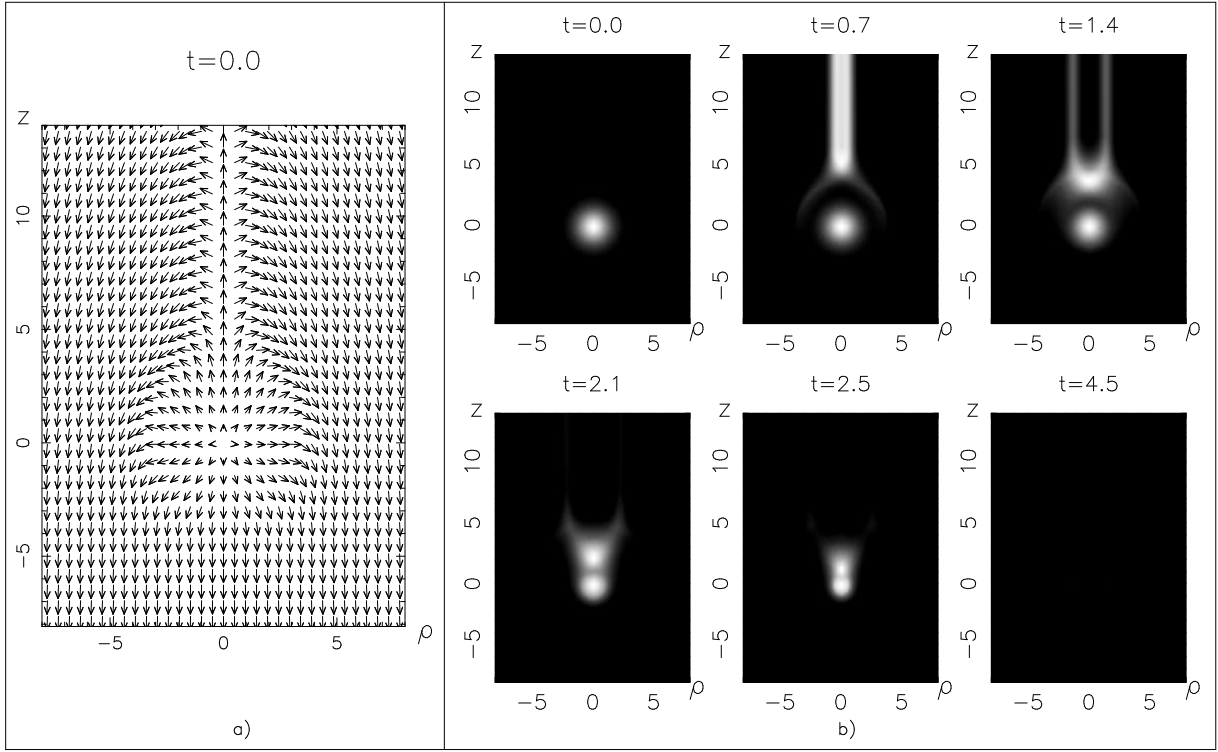}\\
\caption[image]{\label{image} a) Plot of the scalar field $\Phi^a$ in
the configuration given by eq. (\ref{init}) in the text, for $r_1=3$,
$r_2=6$, $a=0.9$, $b=1$. The configuration is axially symmetric.  b)
The result of numerical integration using a) as initial condition.
Potential energy is shown in greyscale at different times. After
annihilation of the monopole-antimonopole pair near $z=0$, the system
fluctuates for some time until all the energy is radiated away (not
shown; note the longer time elapsed between $t=2.5$ and $t=4.5$). }
\end{figure*} 
As explained in \cite{BR,P}, the shift (\ref{globalshift})
  creates
a tension pulling the monopole core, and the apparent unwinding 
(which starts in the inner shells) is only a manifestation of the core's 
translation.
In order to stop the motion of
 the core, we consider a hybrid configuration
such that the monopole core remain unperturbed and the
unwinding occur in the outer shells.  This is achieved by taking
$\xi=0$ for, say, $r<r_1$, a string-like configuration for $r> r_2$,
and some continuous interpolation in between.  One such configuration
(see figure \ref{image}a) would be:
\begin{equation}
{\hat F}(r,y)=f(r), \quad {\hat \xi}(r)=\left\{\begin{array}{ll} 0 &
r<r_1\\ {\it c} \left(1-{{r_1}  \over r}\right)& r_1<r<r_2\\ {\it
a}+{\it b}\ln(r)&r_2<r
\end{array}\right.
\label{init} \end{equation}
with ${\it a}\simgeq {\it a}_0$, ${\it b}\simgeq {\it b}_0$, and $c$ given by 
continuity of ${\hat \xi}(r)$.

If the energy (i.e. mass) inside $r_2$ is large enough there will be no
 appreciable motion
of the monopole core (because the tension of the string is constant 
$\sim 4 \pi$). 
For instance, simulations
with $r_1=3$, $r_2=6$, ${\it b}=1$ show different behaviour depending on
 ${\it c}$: for
${\it c}\simleq 2.8$ the core translates, but for
${\it c}\simgeq 2.8$, the unwinding happens far from the core, which remains
fixed. In the latter case, the decay of the string can also be understood as
a monopole-antimonopole pair creation with the new monopole appearing
at $r=R$ and the antimonopole
appearing near $r=r_2$. 
The equations
solved in the simulations were
\begin{equation} \Box \Phi^a + \Phi^a (|\Phi|^2-1) + \gamma \dot\Phi^a= 0
\quad ,
\end{equation} with a dissipative term added to make the integration
faster. Simulations using different $\gamma$ show no appreciable
difference.  The equations were integrated using cylindrical
coordinates ($\rho,\varphi,z$) in a $200^2$ grid using explicit
Runge-Kutta method with step size control \cite{DoPri}.
The output can be seen in
fig. \ref{image}b, where the potential energy is plotted at various
times, confirming that the configuration (\ref{init}) is unstable and
decays to the vacuum.

  We will now
show that the extra energy required to reach this unstable
configuration (\ref{init}) from the spherical monopole
 configuration (\ref{mon}) is
{\it finite}. 

Consider the set of configurations with $F=f(r)$ and
$\xi=\xi(r)$. From (\ref{energy}) the difference between the energy of
any such configuration ($E[\xi]$) and the energy of the spherically
symmetric unperturbed monopole ($E[0]$) is:
\begin{eqnarray}
E[\xi]-E[0] &=& \left[ \int_{0}^{r_*} +  \int_{r_*}^R \right]\,
\rd r\,\left[ {{r^2} \over 2} f^2(r)
 \xi_r^2(r) I[\xi(r)] \right]\nonumber\\
I[\xi]&=&\int_{-\infty}^{+\infty} \rd y \, \sech^2(y) \, \sech^2(y+\xi(r))\ \ .
\label{ixi}
\end{eqnarray}
$I[\xi]$ is bell-shaped: from a maximum value $I[0]=4/3$,
it rapidly falls to zero for $|\xi| > \xi_* \approx 5$ as $\sim
16(|\xi|-1) exp({-2|\xi|})$. $r_*$ is the radius at which $\xi(r_*)=\xi_*$.
The first integral is clearly finite.
The second can be estimated using the asymptotic form of
$I[\xi]$ and is negligible for $\xi = {\it a}+ \ln r$ even in
 the limit $R\to \infty$, where one gets $ \sim 16\,\pi e^{-{\it a}}
\xi_* e^{-\xi_*}$.

Moreover, there is a continuous path  in
configuration space connecting the configurations (\ref{mon}) and (\ref{init})
such that $E[\xi] -E[0]$ remains finite along the whole path:
first increase $\xi$ in the outer shells $r>r_*$ 
using Goldhaber's deformation ($\xi_r = 0$) 
until $\xi$ reaches $\xi_*$; then adjust the radial dependence
 to match (\ref{init}).

Thus, we have shown that the extra energy required by the monopole
(\ref{mon}) to go over the energy barrier and decay to the
vacuum is finite
even as $R \to \infty$; moreover, since the monopole energy grows with
R, the ratio \hbox{$(E[{\hat\xi}]-E[0]) / E[0]\to 0$} as $R\to
\infty$. 

\section{Stability to small  perturbations with axial symmetry}

We now turn to the classical stability of (\ref{mon}) by considering
small perturbations parametrised by
\begin{eqnarray}
&F = f(r) + \delta(t,r,y)& \\
&\tan ({\thetabar / 2}) = (1 + \xi(t,r,y)) e^y& \qquad . 
\end{eqnarray}   

\begin{figure}[!tb]
\centering
\leavevmode\epsfysize=7cm \epsfbox{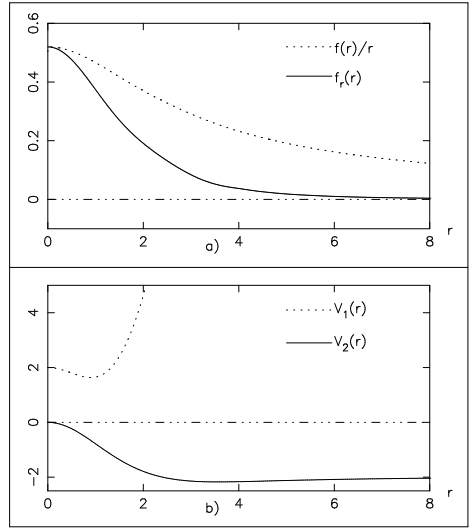}\\
\caption[pot]{\label{pot} The functions: a) $f(r)/r$, $f_r(r)$; b)$V_1(r)$, $V_2(r)$.}
\end{figure}

Neglecting quadratic terms gives $\sin\thetabar \approx \sin \theta +
\xi \sin\theta\cos\theta $, \  $\cos\thetabar \approx \cos \theta - \xi
\sin^2 \theta $. Introducing $X\equiv f\xi$,
\begin{eqnarray}
\Phi^1 &\approx& (f \sin \theta + \delta \sin \theta + X\sin\theta\cos\theta) \cos\varphi \nonumber\\
\Phi^2 &\approx& (f \sin\theta + \delta \sin\theta + X \sin\theta\cos\theta   ) \sin\varphi \nonumber\\
\Phi^3 &\approx& (f \cos\theta  + \delta\cos\theta - X \sin^2\theta ) \qquad ,
\end{eqnarray}
which shows that the correct boundary conditions are that 
$\sin\theta(\delta + X\cos\theta)$ should vanish on the $z$ axis.  Note
 that $\delta(0)$ and $X(0)$ need not vanish. An
infinitesimal translation of the monopole in the $z$ direction
corresponds to $\delta = - f_r(r) \cos\theta, \ X = f(r)/r$, and both $f_r(r)$
and $f(r)/r$ tend to $\approx 0.5$ as $r\to 0$ (see fig \ref{pot}a ).
There is no zero mode
associated with global rotations since these have been factored out in
the ansatz (\ref{ansatz}).

As usual, the perturbation equations

\begin{eqnarray} 
&0 = r^2 \Box\delta +
2\delta + r^2 [3f^2 - 1]\delta +
2 X_y - 4 X \tanh y&  \nonumber\\ 
&0=
\sech^2y \ r^2 \Box  X 
+ \sech^2y [r^2(f^2-1) +2 ] X + 2 \tanh y \, X_y - 2\delta_y &
\label{perteqs1}
\end{eqnarray}                                                          
reduce to an eigenvalue problem in $\omega^2$ for perturbations of the
 form $\delta =
e^{i\omega t} \hat{\delta}(r,y)$, $X= e^{i\omega t}
\hat{X}(r,y)$. Eigenfunctions $\hat{\delta}, \hat{X}$ with negative
eigenvalues $\omega^2 <0$ correspond to instabilities.  Dropping hats and defining $u= \tanh y$
\begin{eqnarray}
&R_1 \delta + \partial_u [(u^2-1)\partial_u \delta]
           -2 \partial_u [(u^2-1) X] = \omega^2 r^2 \delta& \nonumber \\
&R_2 X + \partial_u^2 [(u^2-1) X ] -2\partial_u\delta = \omega^2 r^2 X& ,
\label{perteqs2}\end{eqnarray}
where $R_1$, $R_2$ are radial operators (see fig \ref{pot}b):
\begin{eqnarray}
&R_i = -\partial_r(r^2 \partial_r) + V_i(r),& \qquad i=1,2\nonumber\\
&V_1(r) = r^2(3 f^2(r)-1)+2\nonumber\\
&V_2(r) = r^2(f^2(r)-1)&\qquad .
\end{eqnarray}

Using Legendre polynomials
and changing variables: $\chi := \partial_u [(1-u^2)X] =
\sum \chi_l (r) P_l(u), \ \ \delta = \sum \delta_l(r)P_l(u) $ the equations
 for different values of $l$ decouple, giving
\begin{eqnarray}
&R_1 \Delta_l + x^2 \Delta_l + 2x \chi_l &= \omega^2 r^2 \Delta_l 
\label{eom1}\\
&R_2 \chi_l + x^2 \chi_l + 2 x \Delta_l &= \omega^2 r^2 \chi_l 
\label{eom2}
\end{eqnarray}
where we introduced $x = \sqrt {l(l+1)}$ and
$\Delta_l = x \delta_l$. 
In order to get (\ref{eom2}) we
multiplied the $X$ eqn. in (\ref{perteqs2}) by $(1 -u^2)$ and differentiated
w.r.t. $u$, so there may be spureous solutions; in
particular, $l=0$ corresponds to angular perturbations 
that are singular on the $z$ axis.  These are not physical, and will be
discarded. But
if there is no solution of (\ref{eom1},\ref{eom2}) 
 with negative $\omega^2$
there will be no instability in the original
problem (\ref{perteqs1}).

Our task is to find the solution to (\ref{eom1},\ref{eom2}) with the
minimum value of $\omega^2$ over all admissible perturbations and all
$l \geq 1$. We know one $l=1$ solution, the translational zero
mode ($\tilde{\Delta}_1=\sqrt{2}f_r(r)$, $\tilde{\chi}_1=-2\,f(r)/r$).
Goldhaber's deformation ($\chi_1=f(r)$, $\Delta_1=0$) is also $l=1$
but is {\it not} a solution of (\ref{eom1}, \ref{eom2}), and it can be
shown that there are no instabilities with $\chi_1(r\to\infty)\sim
\rm{const}$.

Let us first consider normalizable perturbations.
Note that, for each $l$, the equations (\ref{eom1},\ref{eom2})
can be obtained by functional variation from
\begin{equation}
E_l \equiv \int \rd r
        \left[r^2\left(\Delta_r^2+\chi_r^2\right)+(V_1+x^2)\Delta^2+(V_2+x^2)\chi^2+ 4x \Delta \chi \right]
   = \omega^2 \int \rd r r^2 [\Delta^2 + \chi^2] \qquad .
\label{el}
\end{equation}

The lowest value of $ \omega^2$ can be found minimising
$E_l$ over normalized functions, $ \int \rd r r^2 [\Delta^2 + \chi^2] = 1 $
and over all $l \geq 1$. However the minimum must be in the
$l=1$ sector since, for all $l>1$ and for given $\Delta$,$\chi$, 
$(E_l - E_1)$ 
is a sum of squares with positive coefficients:                            
\begin{equation}
E_l - E_1= 
  \int \rd r [  A_{l,+} (\Delta + \chi)^2 + A_{l,-} (\Delta - \chi)^2 ]
\ ,
\end{equation}
where
$A_{l, \pm} \equiv  [x^2 - 2 \pm 2(x- \sqrt 2) ] /2 \ >\ 0 $ for $l>1$. 

In order to investigate the $l=1$ sector,
 we rewrite $E_1$ using arbitrary functions $G(r)$, $H(r)$ and $K(r)$:
\begin{eqnarray}
&E_1=\int_0^\infty \rd r\left[\left( r \Delta_r+\frac{G}{r}\Delta\right)^2+
\left(r \chi_r+\frac{H}{r}\chi\right)^2+2 \sqrt{2} \,
\left(\frac{\Delta}{K}+K\chi\right)^2+
\left(V_1+2-\frac{2\sqrt{2}}{K^2}+G_r-\frac{G^2}{r^2}\right)
\chi^2\right.&\nonumber\\
&\left. + \left(V_2+2-2\sqrt{2}K^2+H_r-\frac{H^2}{r^2}\right)\Delta^2\right]
- \left. G(r)\Delta^2(r) \right|_0^\infty 
+ \left. H(r)\chi^2(r)
\right|_0^\infty & \qquad .
\label{big}
\end{eqnarray}
Choosing $G(r)=-r^2\d_r\ln \tilde{\Delta}_1$,
$H(r)=-r^2\d_r\ln\tilde{\chi}_1$,
$K(r)^2=-{\tilde{\Delta}_1}/{\tilde{\chi}_1}$, the coefficients of
$\Delta^2$ and $\chi^2$ in (\ref{big}) vanish identically by virtue of
(\ref{eom}).  For large r, $G$ and $H \sim r$, thus, $E_1 \geq 0$ for
all functions ($\Delta_1$, $\chi_1$) that decay faster than
$1/\sqrt{r}$ as $r\to\infty$. This proves that all normalizable
perturbations have $\omega^2 \geq 0$. The above argument and a host of
numerical simulations strongly suggest that non-normalizable
perturbations are at best compatible with $\omega^2 = 0$, as can be
verified directly from the equations for perturbations
that fall to zero like a power of r, but in this case we have no
analytic proof.

\section{Discussion}

In this paper we have derived and analysed the axial perturbation
equations of $O(3)$ monopoles and proved that, contrary to statements
in the literature, $O(3)$ monopoles are perturbatively stable (or
neutrally stable) to infinitesimal, axially symmetric, normalizable
(or power-law decay) perturbations. We have also proved that the
energy barrier between topological sectors is finite, irrespective of
the details of the scalar potential. This feature is specific of
global monopoles; global vortices in two dimensions, whose energy
grows logarithmically with radius, and gauge monopoles, whose energy
is finite, are separated from the vacuum by an energy barrier growing
linearly with the cut-off. But the energy of global monopoles is
dominated by two-dimensional (angular) gradients far from the core,
and these can be deformed with no energy cost.

One would naively expect thermal fluctuations with $KT
\sim E[\xi]-E[0]$ to cause the monopole to decay.  As far as we know,
this effect is not seen in ``cosmological'' numerical simulations of
global monopole networks, but perhaps this is not so
surprising. First, we worked in flat space.  Second, the range of
scales introduced by the expansion of the Universe forces drastic
approximations on cosmological simulations; in particular, the sigma
model approximation, which is widely used, sets the field on the
vacuum manifold everywhere, so unwinding and pair creation events are
not resolved by the grid.  Finally, cosmological simulations do not
include thermal effects, and it has been shown that full thermal
simulation across the phase transition \cite{AB} gives qualitatively
different results.  Our results provide further evidence that a more
careful analysis of global monopole networks may be required.

\bigskip\hrule\bigskip

\section{Acknowledgments}
 We thank M. J. Esteban for pointing out an error in an earlier 
version of this paper, and  M. Escobedo, R. Durrer, R. Gregory,
 A. Vilenkin, 
K. Kuijken, I.L. Egusquiza, J.M. Aguirregabiria, L. Perivolaropoulos
 and A. Lande for conversations. We acknowledge support from grants CICYT
 AEN99-0315 and UPV 
063.310-EB225/95.


%
\end{document}